# Scintillation muon telescope module with fiber-optic light collection

**Belov**  **S.M., Yanke**  **V.G.**

*The Pushkov Institute of Terrestrial Magnetism, Ionosphere and Radiowave Propagation of the Russian Academy of Sciences (IZMIRAN), Moscow, RF*     yanke@izmiran.ru, sfrovis@gmail.com

**Abstract**

A scintillation muon telescope module with fiber-optic light collection using silicon photomultipliers was developed, tested, and installed for continuous monitoring to study cosmic ray variations. **The aim** of this study was to create a scintillation muon telescope module, continuously monitor the muon component in test mode, and study the long-term stability of the detector parameters. **Methods** for processing the obtained data were developed. To assess the stability of the module parameters, an internal control technique was used, involving data from other detectors. **The results** of testing and long-term continuous monitoring showed that the stability of the developed muon detector is better than 0.1%/year, without the need for its operation in a thermostatic chamber. The study **concludes** that ease of operation, cost, compactness, low power consumption and stability are factors that determine the advantages of the developed module, which is an essential element for constructing a multidirectional muon telescope.

**1. Introduction**

Cosmic rays, which are modulated by the solar wind as they pass through the heliosphere, offer a unique opportunity to study the dynamics of heliospheric disturbances. Given that cosmic rays take minutes to pass through the heliosphere, while heliospheric disturbances take days, continuous recording of cosmic ray fluxes allows us to predict when a disturbance will reach Earth. This approach is used in space weather forecasting and is instrumentally based on the worldwide ground-based Network of neutron monitors and muon telescopes.

Regular ground–based measurements of the intensity of charged particles and muons began in the 1930s using small Geiger-Müller counters and ionization chambers [Forbush, 1958; Shafer et al., 1984].

The era of telescopes began in the early 1930s, when B. Rossi [Rossi, 1966] developed a special coincidence radio engineering scheme. The simultaneous recording of the vertical and directional intensities of cosmic rays by the same set of counters was first achieved in the works of [Kolhorster, 1941; Alfven et al., 1943]. Muon telescopes have evolved: from a single muon detector and an elementary muon telescope to a multidirectional muon telescope [Malmfors, 1948; Elliot, 1949] and a muon hodoscope [Borog et al., 1995].

The first multidirectional muon telescope was not built overnight; the concept of a multidirectional muon telescope evolved over time: from crossed telescopes [Malmfors, 1948; Elliot, 1949] to a multidirectional muon telescope for continuous muon flux recording at Nagoya [Fujii, 1971; Nagashima et al., 1972].

The next stage of the muon telescopes development was the creation of counter muon hodoscopes with a sufficiently high angular resolution of ~7°. This was the muon hodoscope at Norikura [Ohashi et al., 1997]. It was constructed using a matrix type and recorded muons from 1681 directions. At the same time, the TEMP muon hodoscope was created [Borog et al., 1995] based on scintillation strips with a high angular resolution of ~2°. These detectors laid the foundation for the development of muon diagnostics (muonography), which is a set of methods for detailed remote monitoring of various disturbances in the surrounding space, as well as atmospheric and heliospheric processes [Borog et al., 1995; Barbashina et al., 2008].

Not only the geometry of telescopes, but also the detectors that underlie them and record the passage of charged particles, have undergone significant evolution: from gas-discharge counters to scintillation detectors of various types. It is important that instead of collecting light in the form of a truncated receiving cone with the light directed to the cathode of a photomultiplier tube or collecting it using a collecting light guide in the form of a Winston cone, scintillators with fiber optic light collection on an avalanche silicon photomultiplier tube turned out to be effective.

The choice of telescope type is determined by the specific tasks to be solved. In recent years, muon telescopes using modern components have been developed for various scientific and applied applications. For example, [Balabin et al., 2020] described a detector in a telescopic configuration based on scintillators with fiber-optic light collection. The small size of the detectors ensures sufficient statistical accuracy of measurements with 5-minute averaging. [Gerasimova et al., 2021] described an 8 $m^2$ muon telescope with fiber-optic light collection, but with conversion using vacuum photomultiplier tubes.

The requirements for solving cosmophysical problems (geometric factor, angular resolution, optimal number of particle detection directions, long-term stability) can be formulated as follows.

First, achieving the required statistical accuracy of better than 0.1%/hour requires detectors with a total area of several tens of square meters, as the measured charged particle radiation at sea level is ~112 $(m^2 \text{ s sr})^{-1}$ [Grieder, 2001] and is essentially background radiation. Therefore, multidirectional muon telescopes with a high geometric factor $G>2$ $m^2 sr$ are constructed from dozens of identical modules, which are described in this paper.

Second, the main characteristic of a telescope is its angular resolution. It can vary from a few degrees (hodoscopes) to several tens of degrees. For a detailed study of muon flux characteristics and their variations caused by various processes in both the heliosphere and the Earth's atmosphere, detectors with a high angular resolution (~1°) are required. Muon hodoscopes possess these characteristics, the main difference between which and multidirectional muon telescopes lies in the ability to reconstruct the track of each detected muon. If such high spatial and angular resolution is not required, multidirectional muon telescopes are widely used, which are characterized by an angular resolution (~10° or more), which is determined by their design. Heliospheric disturbances propagating from the Sun and modulating cosmic rays are observed from Earth at an angle of several tens of degrees. Indeed, the Larmor radius of protons is $\rho = R/(45B)$, and for an interplanetary field of 5–10 nT and an effective detector rigidity of $R_{eff}$ = 10–100 GV, it reaches 0.2 AU. The angular aperture of the disturbed region near the Sun with these parameters is 20º. This means that detectors on Earth are registering particles collected from a large volume.

Third, the type of detectors and the formation of the electrical signal at the output are important. This determines the complexity, cost, and stability of operation to ensure continuous operation during solar activity cycles. In terms of long-term stability, counter telescopes are the best; they last forever, but large counters are not currently produced. Scintillators with fiber-optic light collection on an avalanche silicon photomultiplier tube have proven very effective.

Fourth, for space weather applications, a multidirectional muon telescope with a total area of at least 4 $m^2$, consisting of $k_X \times k_Y = 4 \times 4$ modules and providing 0.1%/hour for the vertical direction of particle detection, is optimal. Using these "elementary telescopes," it is possible to isolate $(2k_X -1) \times (2k_Y -1)$ independent particle arrival directions with a low angular resolution of ~15°.

Fifth, for our purposes, measurements in the simplest counting mode are sufficient. Other characteristics, such as particle type identification, energy measurement, particle arrival angle measurement, and energy and angular resolution, are unimportant for solar-terrestrial physics applications.

However, scintillation detectors have one drawback: aging, at a rate of approximately 1–2% per year [Kharzheev, 2019]. In [Evdokimov, 2024], the long-term stability of scintillation detector characteristics was studied using a direct method and an accelerated aging method (at high temperature). It was found that, initially, aging proceeds exponentially (1.5% per year) and reaches a plateau with a slope of ~0.5% per year, consistent with aging theory [Zolochevsky et al., 2008]. Further work on assessing long-term stability using direct methods is planned after a sufficient experimental base has been established.

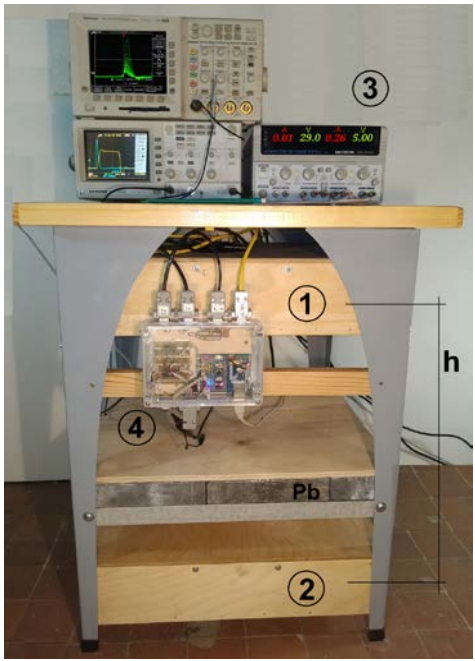

**Figure 1.** Muon telescope module with fiber-optic data acquisition: 1, 2-upper and lower scintillator with silicon SiPM photomultiplier, 3-GPS-74303 power source, 4-coincidence and data acquisition system.

The goal of this work is to create a scintillation muon telescope module using modern components as an element of a multidirectional muon telescope. To achieve this goal, the following tasks must be completed: 1) development and construction of a scintillation muon telescope module with fiber-optic light collection; 2) evaluation of stability to ensure long-term monitoring of the secondary cosmic rays muon component; 3) construction of a data collection and processing system; 4) development of a data correction method for meteorological effects (barometric and temperature). The novelty of this work lies in the implementation of a real-time mode at all stages of data collection and processing, using an atmospheric model in reanalysis mode to account for the temperature effect of the muon component in real time.

## 2. Scintillation telescope module composition

The telescope module (Figure 1) is a particle detector in a telescopic configuration based on plastic polystyrene scintillators. The scintillators were manufactured at the Institute of High Energy Physics (Protvino) [Gorin et al., 2015]. The telescope module consists of two layers of $0.5 \times 0.5 \times 0.05$ m$^3$ scintillators, packed in wooden boxes (Figure 1, blocks #1 and #2), spaced 0.5 m apart. To suppress the soft component of cosmic rays, a lead absorber with a thickness of h = 5 cm is placed between the layers. This geometry forms a cubic telescope, limiting the solid angle of particle reception to 54.7°.

For fiber optic light collection, two 2×2.6 m long optical fibers are glued into the scintillator. The scintillator has a maximum emission spectrum of 425 nm with a fiber converter that shifts the wavelength to 476 nm. The fiber ends are connected to the corner point of the scintillator.

A SensL MicroFC-30035-SMT avalanche silicon photomultiplier (SiPM) is used as the photodetector [Gorin et al., 2015]. The gain is at least $3 \cdot 10^6$ and the operating bias voltage is 29 V. Silicon photomultipliers have the highest sensitivity to blue light and cover the spectral range of 300–950 nm. The matrix contains 4774 cells; the active region of the matrix is $3 \times 3$ mm$^2$. The small dimensions of the SiPM and preamplifier allow the creation of detectors without ineffective geometric zones and their use in forming detectors of any size, which is important when designing multidirectional muon detectors.

Silicon photomultipliers are a modern alternative to vacuum photomultipliers. The main disadvantage of silicon photomultipliers is their relatively high noise level at temperatures above room temperature. However, the sensitivity's dependence on temperature can be taken into account when processing the data.

## 3. Telescope module amplifier

The signal from the MicroFC-30035-SMT silicon photomultiplier of each telescope layer is fed to an amplifier-discriminator and shaper [Gorin et al., 2015], integrated into the scintillator module (Figure 2). The output signal from the silicon photomultiplier is amplified by a factor of $3 \cdot 10^6$ (amplifier DA1). The pulse duration is approximately 100 ns, with edges lasting a few nanoseconds. The signal is shaped by a threshold set by potentiometer R8 (analog comparator D1) and is shaped by a duration of 400 ns (pulse shaper D2). The typical discrimination threshold is 1 volt.

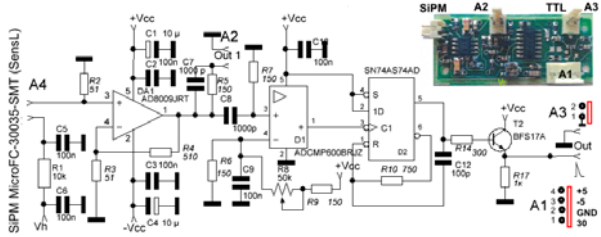

**Figure 2.** Discriminator amplifier and pulse shaper for SiPM with TTL output. Insert – external view of the amplifier.

Compared to our previous prototype, the new electronics demonstrate higher performance, greater efficiency, and lower power consumption, and also features a new data acquisition system with an increased dynamic range.

## 4. Telescope data acquisition system

The recording system is based on an STM32f103c8t6 (ARM Cortex-M3) microcontroller (Figure 1, block #3). Positive-polarity output pulses generated on the amplifier-discriminator board, each 400 ns in duration, are fed to a coincidence circuit. Signals from the upper and lower detectors, as well as from the coincidence circuit, are fed to the controller input, which processes the interrupt for each pulse and counts them over a set 60-second acquisition time. The processing time for each interrupt is 800 ns. Since the load on each channel does not exceed 100 pps, the number of counts will be no more than $10^{-4}$ %.

In addition to the telescope's count rate, the data acquisition time also records atmospheric pressure and ambient temperature (BMP280 sensor, ±0.12 hPa, ±0.01°C) or a standalone digital thermometer (DS1631AU, ±0.5°C at –55 ÷ 125°C), a remote temperature sensor (DS18B20 sensor, ±0.5°C), and the bias voltage (±0.001V) of the silicon photodetector using the microcontroller's 12-bit integrated ADC.

To maintain accurate time, a DS3231 real-time clock is used, periodically synchronizing with the global time over the network using the NTP protocol, ensuring a clock accuracy of ~±5 seconds per month.

Data is transferred to the server over a TCP/IP network using a W5500 Ethernet controller. Data is sent every minute. In case of problems with sending data to the server, the system provides the option of autonomous operation with data storage for up to 10 days in the non-volatile flash memory AT25DF321 with a capacity of 32 Mbit, followed by sending it to the server.

Data received by the server undergoes initial processing, but the software package is designed in such way that additional data (meteorological model) collection and processing occurs exclusively upon a corresponding client request. This solution increases the average latency for results, but avoids unnecessary calculations and reduces the load on both the main system server and the data source server. Following a client request, the server queries the meteorological or forecast model data. Data from the corresponding detector or GSM model results are also retrieved to account for primary variations. The next step is to correct the initial data using the selected method.

The software package implements a classic three-tier architecture. The server part is a complete software package implementing the logic for receiving and processing muon detector data, including atmospheric pressure, the logic for obtaining data on the vertical temperature variation in the atmosphere above the location point, data correction, and other service functions. The software suite was developed using Python 3.9.0 for server-side calculations and data processing, and JavaScript for the interface and graphical presentation on the client. PostgreSQL, an open-source database management system, is used to store final and intermediate data processing results.

The software package allows for querying muon detector and meteorological data, correcting the data for meteorological effects, and determining (upon request) the barometric and temperature coefficients of the cosmic ray muon component.

A web browser is used as the software suite's thin client. This, along with the choice of client-server architecture, significantly facilitates user access to monitoring results. The interface utilizes modern web technologies for ease of use. The interface's primary functionality is a graphical representation of calculation results for the requested period: https://tools.izmiran.ru/w/muon (select Moscow-pioneer, for example).

## 5. Telescope module characteristics.

*5.1. Detector Calibration.* The SiPM breakdown voltage was set for each channel. A precision power supply (GPS-74303 (0.01% instability) or PWS-4305/PWS-2721) with built-in temperature compensation is used for power supply. A local temperature and bias voltage sensor for the solid-state silicon photomultiplier is also used for monitoring.

The detector's operation at bias voltage

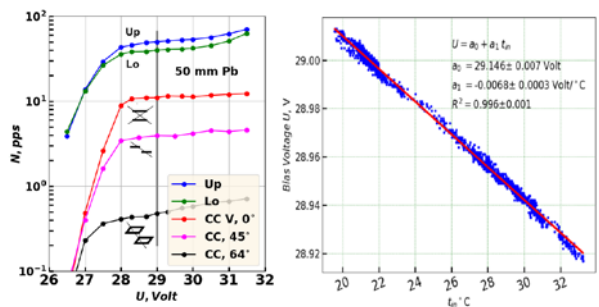

**Figure 3.** Dependence of the count rate on the bias voltage on the SiPM for the upper and lower detectors, for coincidences in cubic (0°), inclined (45°) and azimuthal (64°) geometries (Pb thickness 50 mm).

U is determined by the slope of the muon counting efficiency plateau.

The nominal operating voltage of the SiPM is 29.0±0.1 V. The measured counting curves (Figure 3, right panel) were used to estimate the gain dependence on supply voltage and the SiPM power supply requirements. Analysis showed that the counting curve slope at 29 V for single channels is 0.01%/mV, while for the coincidence channel; it is 0.003%/mV for the vertical. The statistical accuracy of the count rate with hourly averaging is 0.3%. Therefore, the requirement for bias voltage stability is no worse than 0.1 V. The sources used provide such stability.

The bias voltage dependence on $t_{in}$ (time) is fitted with the function $U = 29.146 - 0.007\ t_{in}$ (V). The actual annual temperature variation is 10 °C, therefore $\Delta U = 0.01$ V. The count curve is monitored in field conditions, and a muon momentum spectrum is also plotted in the laboratory to select the operating point.

*5.2. Detector Efficiency.* The efficiency $\varepsilon_x$ of the scintillation detectors was estimated by the manufacturer [Gorin, 2015] using a standard method, using two $N_1$ and $N_2$ test scintillation detectors measuring 20×20×5 cm³, which were located above and below the X detector under study. The detector counting efficiency was calculated as the ratio of the number of triple coincidences $N_{1X2}$ (1st test $N_1$, test $N_x$, and 2nd test $N_2$) to the two-fold coincidence of the detectors $N_{12}$ with the same detector geometry and subtracting false coincidences (see 5.3). The efficiency in this case is $\varepsilon_X = N_{1X2}/N_{12}$ and for our detectors is equal to (98.7±0.2)%.

The optical characteristics of scintillators deteriorate under the influence of background radiation, or they degrade naturally under the influence of ambient temperature and humidity, leading to the formation of microcracks on the surface and inside the scintillator. Under normal temperature and humidity, the loss in light yield is approximately 1–2%/year [Kharzheev, 2019; Zhang et al., 2023; Zolochevsky et al., 2008], which requires correction when studying long-term variations of cosmic rays. An approximate correction can be made based on the presented data, but it is preferable to use a "reference" detector. A suitable "reference" detector is a proportional counter detector, which has worse statistics but significantly better long-term stability due to the absence of degradation of its individual elements.

*5.3. Miscalculations and Random Coincidences.* Due to the relatively short dead time and low detector load, the number of miscalculations associated with the dead time $\tau$ and the number of random coincidences are negligible.

Indeed, with a decreasing particle number $N_0$, the number of counts is $\tau N_0$. The measured number of particles $N$ is then defined as $N=N_0/(1+\tau N_0)$. The variation in the number of counts associated with the change in dead time is $\delta N/N = -\delta\tau\, N$. For a dead time of $\tau = 400$ ns, a change in dead time $\delta\tau$ of no more than 50 ns, and a decreasing particle number $N_0 = 50$ sec$^{-1}$ (single counter), the number of counts is $\delta\tau N_0 = 1\%$, and the variation in the number of counts is 1%.

The total number of multiple random coincidences is defined as $k\tau^{k-1}N^k$. Then the number of random double coincidences is 0.02%, and the variation in the number of random double coincidences associated with the change in dead time is $\delta N/N = 2\delta\tau\, N = 1\%$. In this case, the number of random double

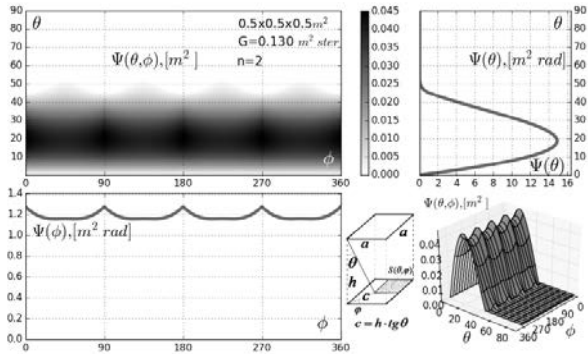

**Figure 4.** Integral, zenith and azimuthal beam patterns for recording the muon intensity of a cubic telescope.

coincidences is negligible compared to the useful signal.

*5.4. Geometrical characteristics of the detector.* The directivity pattern and the geometric factor in the approximation of the angular dependence of the muon intensity are given in analytical form in (Appendix A). The numerical result for the vertical intensity values Io and the geometric factors are given in Table 1, and the directivity patterns for recording the muon intensity of the cubic telescope are shown in Figure 4.

*5.5. Detector Energy Characteristics.* The detector output data are the signals from the upper and lower scintillators and the coincidences between them. At sea level, the intensities of muons, electrons, positrons, soft muons, protons, and photons are approximately 90, 30, 30, 1, and 90 1/(m$^2$ s sr), respectively [Hayakawa, 1973]. The contributions of different types of particles with energies >21 MeV to the upper scintillator count rate in 2π geometry are then approximately 60%, 20%, 18%, 1%, and 1% for μ, $e^\pm$, slow muons μ$_s$, $p$, and γ (with a γ-quanta detection efficiency of 1%). The average count rates measured by the three detectors and the corresponding statistical uncertainties are given in Table 1.

The energy threshold is determined by calculating the amount of matter located directly above the detector. The threshold for detecting charged particles for the upper scintillator is 21 MeV and is determined by losses in the upper scintillator's protective housing and the scintillator itself. The energy threshold for the lower scintillator is <163 MeV and is determined by losses in the two protective scintillator housings, the two scintillators, the lead (5 cm) absorber, and the uncertainty in the electron flux from bottom to top. The telescope's coincidence network limits the particle energy threshold to 163 MeV. Taking into account the muon angular distribution of $\sim\cos^2\theta$, the average energy loss is ~220 MeV. The threshold values are listed in Table 1.

We also present the energy characteristics of the vertical muon telescope at the Moscow latitude for a power-law spectrum of variations (γ=0.7): the mean, median, and effective rigidities of the recorded muon component are 110, 26, and 39 GV, respectively.

For a known count rate $N$, geometric factor $G$, and detector efficiency of the telescope (ε≈98.7), the vertical intensity is determined as [(m$^2$ c sr)$^{-1}$].

*5.6. Meteorological effects.* When monitoring cosmic radiation, variations of various types are simultaneously observed: atmospheric, magnetospheric, and primary. The amplitudes and phases of these variations generally do not differ significantly. Therefore, when studying one type of variation, variations of other types must be excluded. The method for excluding the meteorological effect (barometric and temperature) is considered in the works [Kobelev et al., 2024; Berkova et al., 2018]. The vertical intensities of the muon and total charged components are also given in Table 1.

**Table 1.** Selected characteristics of the MT OPTO telescope.

| | $N$, pps | $\sigma$, % | $> E_x$, MeV | $n$ $\cos^n\theta$ | $G$, $m^2$ sr | $I_0$, $(m^2 \, s \, sr)^{-1}$ | $\beta \pm 0.02$, %/hPa | $\alpha_T \pm 0.04$, %/°C | $\delta \pm 0.14$, %/% |
|---|---|---|---|---|---|---|---|---|---|
| Up | 47 | 0.25 | 21 | 0.6 | 0.605 | 76.9 | 0.22 | 0.22 | 0.69 |
| V | 12 | 0.5 | 220 | 2 | 0.130 | 92.3 | 0.17 | 0.29 | 0.64 |
| Lo | 40 | 0.25 | <163 | 0.6 | 0.605 | ~65.3 | 0.19 | 0.27 | 0.53 |

To estimate the coefficients of the barometric $\beta > 0$ and temperature $\varepsilon > 0$ effects, a model of atmospheric variations was constructed

$$N_C = N_U \exp[-\beta(P_0 - P)] \times \\ \times [1 - \varepsilon(T_{0m} - T_m)] \times (1 - \delta\nu_S). \quad (1)$$

According to this theory, to estimate the corrected count rate $N_C$, the measured detector count rate $N_U$ must be converted to the mean atmospheric pressure $P_0$ and the mass-average temperature $T_{0m}$, $P$ and $T_m$ are the current atmospheric pressure and mass-average temperature. The results of the multivariate

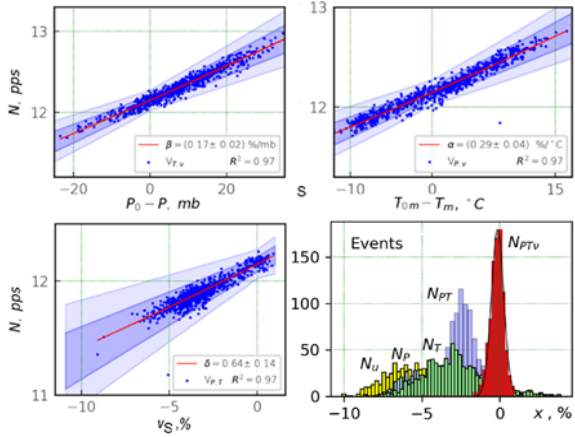

**Figure 5.** Scatter plot illustrating the correlation between the vertical muon telescope count rate and pressure (upper left panel), the mass-averaged atmospheric temperature (upper right panel), and primary variations (lower right panel). The regression line and 1σ and 2σ error bars are shown for each case. The lower right panel shows the distributions after each correction step.

correlation analysis (1) are presented in Table 1 and Figure 5.

Table 1 and Figure 5 present the linear regression coefficients with errors and the multiple determination coefficients $R^2$. The $R^2$ values indicate the quality of the model and the adequacy of the formulated mathematical model (1). Indeed, the value of the multiple determination coefficients $R^2$ means that 97% of the variations are explained by the constructed regression equation (1).

The lower right panel Figure 5 shows the distribution of the original $N_U$ signal and the distribution after correction for pressure, temperature, and primary variations. The normal distribution of NPTv indicates the adequacy of model (1).

The easily accounted for local barometric effect requires only precise atmospheric pressure at the observation point [Kobelev et al., 2024]. Estimating the distributed temperature effect requires knowledge of the altitude temperature distribution. As the atmospheric temperature distribution increases, the geometric thickness of the atmosphere increases, which at the Earth's surface leads to a decrease in muon intensity due to their decay.

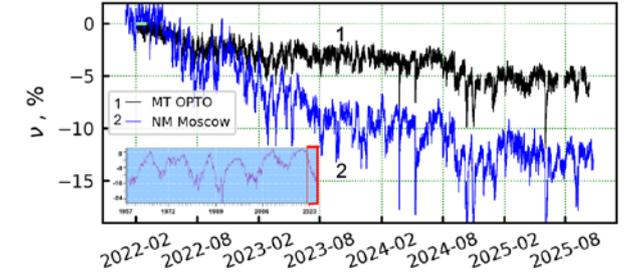

**Figure 6.** Comparison of count rate variations between the MT OPTO muon telescope module and the Moscow neutron monitor. The tab shows variations of the neutron component in Moscow relative to the 1987 baseline.

## 6. Continuous monitoring results

The telescope has been operating in test mode since October 2021, and has been in continuous recording mode in Moscow since February 5, 2022. The detector's coordinates are 55.47°N and 37.32°E, its altitude is 200 m, and its vertical geomagnetic cutoff rigidity is 2.09 GV (for 2020). Real-time muon telescope data (upper and lower detector count rates and the vertical hard muon telescope with minute and hourly resolution) are available at https://tools.izmiran.ru/w/muon (select Moscow-pioneer). The data has been corrected for barometric and temperature effects.

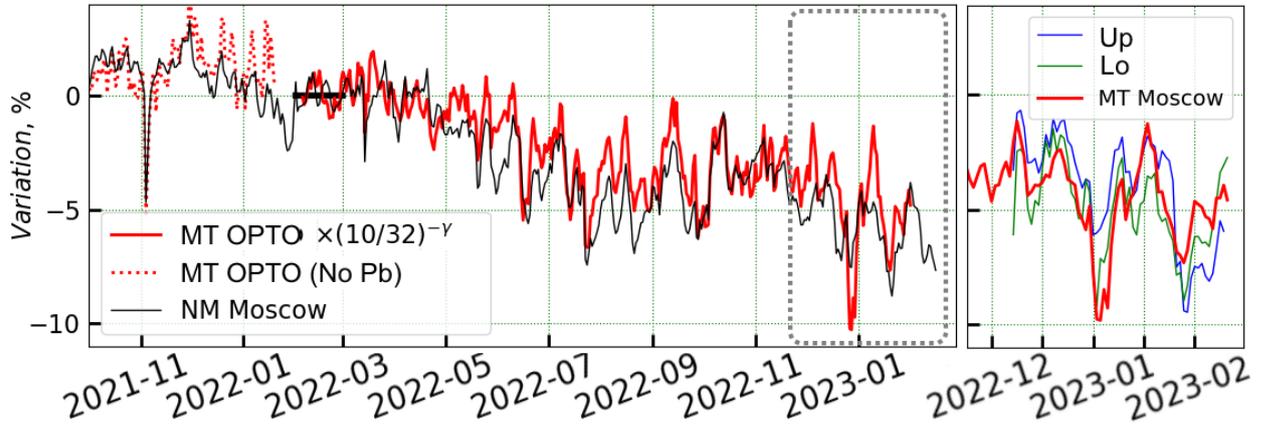

**Figure 7.** Comparison of the count rate variations of the Moscow neutron monitor and the count rate variations of the MT OPTO muon telescope module normalized to 10 *GV*.

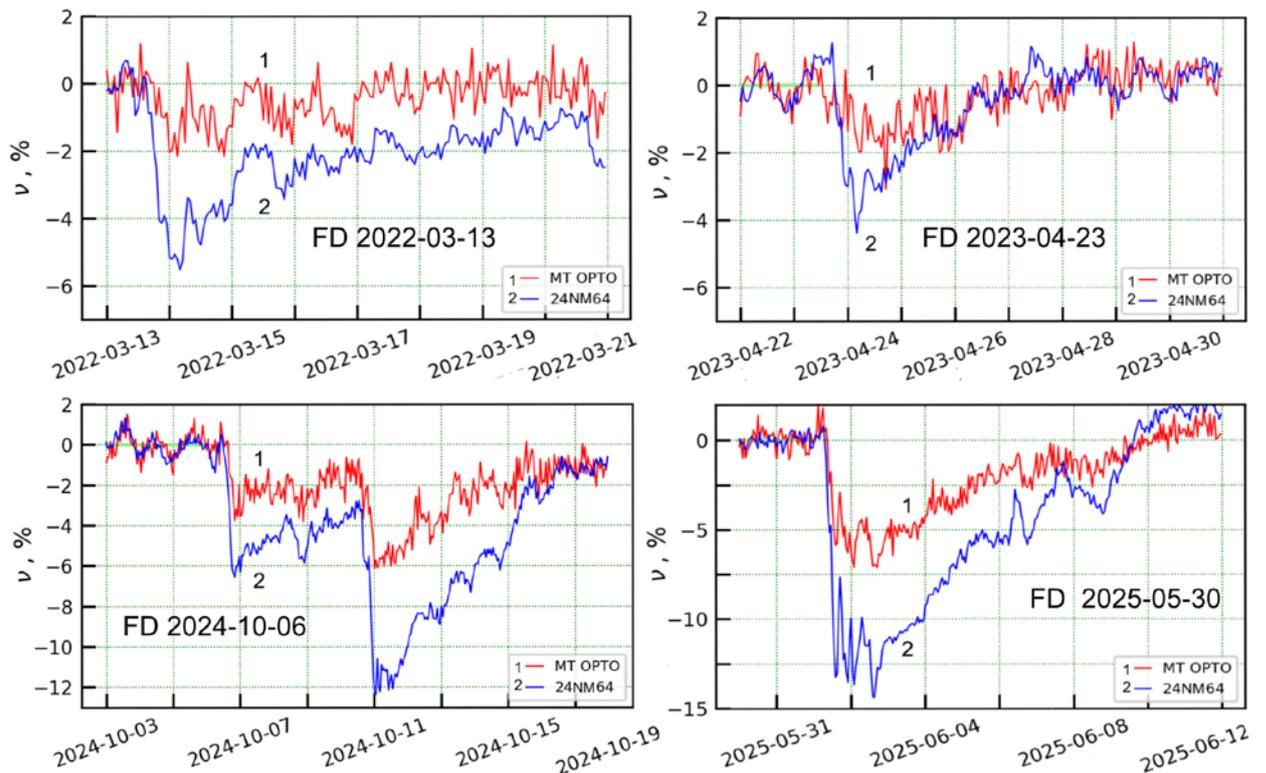

**Figure 8.** Forbush decreases on March 13, 2022, April 23, 2023, October 6, and May 30, 2025, based on hourly resolution neutron and muon components in Moscow.

Figure 6 shows the results of recording variations in the vertical count rate of the MT OPTO telescope and the Moscow neutron monitor for the entire observing period (baseline period – February 2022). The effective rigidity of the recorded particles for the power-law spectrum of long-term variations ($\gamma = 1$) is $R_{eff} \approx 32$ GV for MT and $R_{eff} \approx 10$ GV for NM.

In Figure 6, the different time courses of the detector count rates are due to the different effective rigidities of the detecting particles: the neutron monitor is ~10 GV, and the muon telescope is ~32 GV. The variation in the muon component normalized from an effective rigidity of 32 GV to 10 GV is compared with the variations in the neutron component in Figure 7. During the tuning phase, there was no lead absorber between the scintillator layers (Figure 7, dotted line). A consistent time course is evident.

Continuous observations were conducted near the SA maximum, when approximately 880 Forbush decreases were observed since the beginning of 2022, 41 of which were large: from 3% to 15%. The most interesting events, one for each year, are compared in Figure 8 for the muon telescope and the neutron monitor.

Figure 8 (upper left panel) shows the Forbush decrease for March 2022. It is associated with intense flares on March 13, which also caused disturbances in the Earth's magnetosphere. These disturbances lead to a decrease of the geomagnetic cutoff rigidity, especially for mid-latitude detectors, which affects the PE profile. However, the disturbance manifests itself not as a decrease in CR intensity, but rather as a slight increase (in the figure, this is the end of March 13).

The Forbush decrease in April 2023 (upper right panel) is a decrease with a classic profile and a slight magnetospheric effect around April 23-24, 2023.

Figure 8 (lower left panel) shows a double Forbush decrease in October 2024, while the lower right panel shows the largest Forbush decrease in May 2025 over the last two solar cycles.

### 7. Detector data availability

Real-time muon telescope data is available at https://tools.izmiran.ru/w/muon (application → "Moscow-pioneer"). Various characteristics (coupling functions, reception coefficients, temperature coefficient densities) can be found in the "MuonTelescope" folder on the Yandex Disk resource [https://disk.yandex.ru/d/mKHMM2dztqNoHw].

### 8. Conclusion

A scintillation muon telescope module has been developed using modern components as an element of a multidirectional muon telescope.

**1)** The developed telescope module has several key features and is designed for monitoring muon fluxes in cubic or semi-cubic geometry. The module's design allows for the construction of multidirectional telescopes required for solving specific physics problems, as well as for use as a simple muon telescope. The two detector layers, separated by a lead absorber layer, enable simultaneous measurement of both the hard component of secondary cosmic radiation (muons) and the soft component of the carpet detector in $2\pi$ geometry (soft muons, electrons, and positrons). The threshold value for measured muons is 220 MeV. The geometric factor of the cubic telescope is $S\Omega=0.13$ m² sr, and that of the carpet detector is approximately 0.60 m² sr. The beamwidth at half maximum for the cube telescope is 30°.

**2)** The required bias voltage stability of the silicon photomultiplier (no worse than 0.1 V) is easily achieved with modern power supplies.

**3)** The constructed data acquisition and processing system ensures almost 100% efficiency.

**4)** Analysis of the obtained data allowed us to estimate with good statistical accuracy the barometric and temperature effects of the muon component, the coefficients of which are $\beta = (0.17 \pm 0.02)$ %/hPa and $\alpha = (0.29 \pm 0.03)$ %/°K, respectively. For the upper detector (the carpet), $\beta = (0.23 \pm 0.05)$ %/hPa and $\alpha = (0.22 \pm 0.07)$ %/°K.

The novelty of this work lies in the implementation of a real-time mode at all stages of data collection and processing, using an atmospheric model in reanalysis mode to account for the temperature effect of the muon component in real time. The results of this work will be applied to data processing from the existing network of muon telescopes. In the future, as data accumulates, it will be important to experimentally evaluate the degradation of efficiency under natural conditions under the influence of ambient temperature and humidity.

## Acknowledgements

The authors are grateful to the participants of the NMDB project (www.nmdb.eu). This work is being conducted within the framework of the Russian National ground-based Network of CR Stations (CR Network) (https://ckp-rf.ru/catalog/usu/433536) and the Ministry of Science and Higher Education's (Ministry of Science) Marine Scientific Research project for 2025 and 2028.

## APPENDIX A: Directional Pattern and Geometrical Factor

For the angular dependence of the detected particles at the observation level $h$, the approximation $I(h,\theta) = I_0(h) Cos^2\theta$ is applicable. Then, the measured count rate $N(h)$ [pps] is related to the vertical intensity $I_0$ [1/(m$^2$ c ster)] and the detector geometrical factor $G(h)$:

$$N(h) = I_0(h)\varepsilon \cdot \int_0^{2\pi} d\varphi \int_0^{\pi/2} \sin\theta d\theta \cdot S_\perp(\theta,\varphi) Cos^2\theta = I_0(h)\varepsilon \int_0^{\pi/2}\int_0^{2\pi} \Psi(\theta,\varphi) d\theta d\varphi = I_0 \varepsilon \cdot G(h), \quad (A.1)$$

where $\varepsilon = \varepsilon_{Up}\varepsilon_{Lo}$ is the product of the efficiencies of the upper and lower detectors, and $S_\perp(\theta,\varphi) = S(\theta,\varphi)\cos\theta$ is the area intersected by the particles. The thickness of the detectors was neglected.

The radiation pattern $\Psi(\theta,\varphi)$ and geometric factor $G(h)$ of the detector are respectively defined as:

$$\Psi(\theta,\varphi) = Sin\theta \cdot S(\theta,\varphi) \cdot Cos\theta \cdot Cos^2\theta \ [m^2] \quad (A.2)$$

$$G(h) = \int_0^{\pi/2}\int_0^{2\pi} \Psi(\theta,\varphi) d\theta d\varphi = \int_0^{\pi/2} \Psi(\theta) d\theta \ [m^2 \ sr] \quad (A.3)$$

Following the radiation pattern calculation methodology of [Sullivan, 1971; Thomas & Willis, 1972; Pak, 2018], we present here the result for our geometry, namely, for a vertical telescope and for a carpet detector. We used the "shadow area" method proposed in these works, which, unlike the "infinitesimal area" method, allows us to calculate both the radiation patterns and the geometric factor of the detector. Denoting $c = h \cdot tg\theta$ and $\xi = a/h$, the radiation pattern (A.2) can be simplified by transforming $S(\theta,\varphi)$ into the expression

$$S(\theta) = 4 \cdot Sin\theta \cdot Cos^3\theta \begin{cases} \dfrac{\pi}{2}a^2 - 2ac + \dfrac{1}{2}c^2, & 0 \leq \theta \leq arctg\,\xi \\ a^2 ArcSin\dfrac{a}{c} - \dfrac{1}{2}a^2 - ac + a\sqrt{c^2 - a^2}, & arctg\,\xi \leq \theta \leq arctg\sqrt{2}\xi \end{cases} \quad (A.4)$$

An analytical expression can be obtained for the geometric factor

$$G_{n=2} = S\Omega = \dfrac{a^2}{\xi}\left[\dfrac{(1+2\xi^2)}{\sqrt{1+\xi^2}} arctg\dfrac{\xi}{\sqrt{1+\xi^2}} - arctg\,\xi\right] \quad (A.5)$$

For our geometry, $a \times h) = 0.5 \times 0.5$ m$^2$ $G_{n=2} = 0.130$ m$^2$ ster, i.e., the angular aperture $\Omega \approx 0.52$ ster. The inequality $G \leq a^2 \cdot a^2/h^2$ [Sullivan, 1971] is always valid. The beam patterns for recording muon intensity for the cubic telescope are shown in Figure 3.

The geometric factor of the upper and lower single "Carpet" detectors for a given two-dimensional beam pattern is calculated as:

$$G\_(x) = \int_0^{\pi/2}\int_0^{2\pi} \Psi\_(\theta,\varphi) d\varphi d\theta = \int_0^{\pi/2} \sin\theta d\theta \int_0^{2\pi} d\varphi S \cos^{n+1}(\theta) = \dfrac{2\pi}{(n+2)} S. \quad (A.6)$$

The numerical values of the vertical intensities $I_0$ and geometric factors are given in Table 1. The angular zenith distribution index for hard muons is taken to be $n=2$, and for a single flat detector $n=0.6$ (electrons, soft muons, muons). The beamwidth at half maximum for the cubic telescope is 32°.

## APPENDIX B: Polar Coupling Functions.

*Neutron component Coupling Functions.* Neutron coupling functions for latitude measurements are determined experimentally. The most effective approximation of the latitude curve is the Dorman–Granitsky approximation [Dorman et al., 1970] in the form:

$$I(R) = I_0[1 - \exp(-\alpha R^{-(\kappa-1)})], \tag{B.1}$$

where $I_o$ is the count rate at the pole. Polar coupling functions, as is known, can be obtained from the latitude curve of the count rate by differentiating it, i.e. $W(R) = -I^{-1} \partial I/\partial R$, and then

$$W(R) = \alpha(\kappa-1)\exp(-\alpha R^{-(\kappa-1)})R^{-\kappa} \tag{B.2}$$

This approximating function is also remarkable in that the normalization condition is satisfied for any values $\alpha$ and $\kappa$, i.e. $\int_0^\infty W(R)dR = 1$. Differential polar and integral functions of the connection are presented in Figure B.1.

The coupling functions for a station with rigidity $R_c$ can be obtained by renormalization, i.e. $W(R_c, R) = W(R) / \int_{R_c}^\infty W(R)dR$ for rigidity $R \geq R_c$.

Several simple and useful expressions can be written that characterize the coupling functions and are determined only by the values $\alpha$ and $\kappa$.

Табл. B.1

| | |
|---|---|
| $W_{max} = \kappa \exp(-\kappa/(\kappa-1))/R_{max}$ | maximum value of the coupling function |
| $R_{max} = (\alpha(\kappa-1)/\kappa)^{1/(\kappa-1)}$ | rigidity at which it is achieved $W_{max}$ |
| $R_l = 0.1 * R_{max}$ | minimum value of the coupling function on the low rigidity side (at the 0.1 level) |
| $R_{med} = (\alpha/\ln 2)^{1/(\kappa-1)}$ | median rigidity |
| $s = 1 - \exp(-\alpha R_{eq}^{-(\kappa-1)})$ | amplitude of the latitude effect (rigidity at the equator) |

The coupling functions of the neutron component for various observation levels were calculated by solving a system of kinetic equations describing the generation and propagation of the neutron component in the atmosphere. Approximation (B.2) was performed using these results. The altitude dependence of the parameters $\alpha$ and $\kappa$ for the minimum and maximum solar activity can be represented by the following expression (h in bars) [Aleksanyan et al., 1982]:

| | |
|---|---|
| or the minimum (1965) of solar activity (h$_0$ in bars)<br>$\ln\alpha = 1.84 + 0.094 \cdot h_0 - 0.09 \cdot \exp(-11h_0)$, $\kappa = 2.40 - 0.56h_0 + 0.24\exp(-8.8h_0)$ and<br>for the maximum (1969) of solar activity:<br>$\ln\alpha = 1.93 + 0.150 \cdot h_0 - 0.18 \cdot \exp(-10h_0)$ $\kappa = 2.32 - 0.49h_0 + 0.18\exp(-9.5h_0)$ | (B.3) |

Up to a level of 300 hPa, the dependence of the parameters of interest to us on altitude is almost linear.

*Muon component Coupling Functions.* Coupling functions of primary and secondary cosmic ray variations characterize the relative sensitivity of the detector to primary cosmic ray protons of different energies. Coupling functions can only be obtained theoretically by solving a system of kinetic equations describing the generation and propagation of secondary components in the atmosphere. In [Fujimoto 1976; 1977], an approximation of the resulting directional coupling functions of the muon component was performed. It is proposed to approximate the directional unnormalized coupling functions of the muon component, recorded at a zenith angle θ, by an expression of the form:

| | |
|---|---|
| $W(R, h_0, \theta) = \beta \cdot \exp(-\alpha \cdot z^{-\delta}) \cdot z^{-\kappa}$, $z = R/R_m$, где $R_m$ - median rigidity in GV. The approximation parameters $\alpha, \kappa, \beta, \delta$ are defined as follows: $\alpha = 0.37 R_m^{0.1}$, $\kappa = 1.82 R_m^{0.02}$, $\beta = 540 R_m^{-(0.82+0.2\ln R_m)} muon/(m^2 \cdot s \cdot sr \cdot GV)$, $\delta = 1.1$. | (B.4) |

If the muon detector is located at a depth $h_0$ of m w.e. in the atmosphere and under a shield of thickness $d$ m w.e., which cuts off muons with energies (e.g., for $\Delta\varepsilon=0.316$ GeV $d=0.025$ m w.e.), then the median rigidity can be approximated as

$$R_m = 2.25 \left(\frac{\lambda h_0 + d}{\cos\theta}\right)^{1.04} \quad \text{or} \quad R_m = 2.25 \left(\frac{\lambda h_0}{\cos\theta} + d\right)^{1.04} \quad (B.5)$$

in the case of a spherically symmetric screen (e.g., for an ionization chamber). The coefficient $\lambda$ regulates the dependence of the coupling functions on the solar activity level: $\lambda = 2$ and $\lambda = 2 + 1.013/h_0$ for the minimum and maximum solar activity, respectively, which is especially important for the upper atmosphere, such as stratospheric measurements.

The median rigidity for vertical particle arrival is 50.8 GV. The maximum value of the coupling function is achieved at a rigidity of $R_{max} = 0.2562 R_m^{1.0727}$, and in our case, is equal to 17.3 GV.

Табл. B.2

| | |
|---|---|
| $R_m = 2.25 \left(\frac{2h_0 + d}{\cos\theta}\right)^{1.04} = 2.25 D^{1.04}$ | Median rigidity (flat-parallel screen, muon telescope) |
| $R_m = 2.25 \left(\frac{2h_0}{\cos\theta} + d\right)^{1.04} = 2.24 D^{1.04}$ | Median rigidity (spherically symmetric screen, ionization camera) |
| $W_{max} = \beta \exp(-\kappa/\delta)(\frac{\kappa}{\alpha\delta})^{\kappa/\delta} = \beta \exp(-\kappa/\delta)(\frac{R_m}{R_{max}})^{\kappa}$ | Maximum value of the connection function |
| $R_{max} = (\frac{\alpha\delta}{\kappa})^{1/\delta} R_m = 0.2562 R_m^{1.0727}$ | The maximum value of the coupling function is achieved at rigidity |
| $R_L = 0.1288 D^{1.092}$ | Minimum value of the coupling function on the low rigidity side (at the level of 0.01) |

Thus, the coupling function method allows one to take into account the generation of the muon component measured by the detector during the interaction of the primary particle flux with the Earth's atmosphere.

In Table B.1 provides several simple and useful expressions characterizing the coupling functions, which are determined solely by the values of the approximation parameters.

Unlike the neutron component, the coupling functions of the muon component require normalization in all cases, according to the expression $\int W(R, h_0) dR$, since the normalization condition is not automatically satisfied

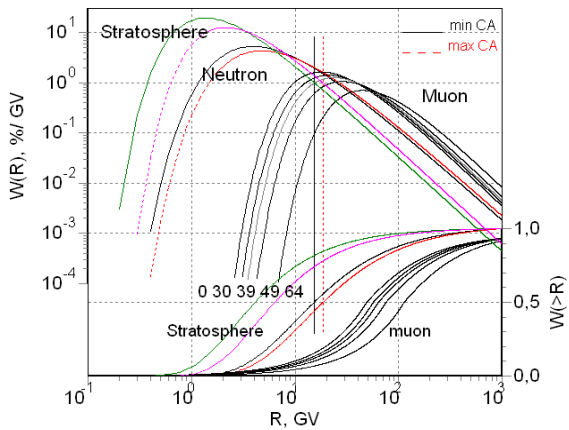

**Figure B.1.** Differential polar and integral coupling functions of the neutron monitor, the Nagoya muon telescope (vertical and inclined) and the charged component telescopes of the stratospheric sounding system.

Figure B.1 compares the polar (differential and integral) coupling functions of the neutron monitor, the Nagoya muon telescope (vertical and inclined), and the charged component telescopes of the stratospheric sounding system. These coupling functions have been tested in solving many cosmophysical problems.

*Stratospheric sounding detector Coupling Functions.* Using stratospheric sounding data allows us to expand the studied energy range to lower rigidities, down to several tenths of a GV. For this purpose, we use data from ionizing radiation counter telescopes near the Pfotzer maximum, whose depth is between 80 and 120 hPa and depends on the solar activity level and the geomagnetic cutoff rigidity. In this case, we can also use an approximation of the coupling functions of the form (18). The coupling function for the 100 hPa level is also shown in the summary figure 2 for the minimum and maximum solar activity. We limited ourselves to three sites—Murmansk, Moscow, and Mirny—for which a long series of observational data is available and where such measurements are currently being conducted.

*Spacecraft detectors Coupling Functions.* Using data from spacecraft detector monitoring allows us to further expand the energy range under study to the region of low rigidities, down to a few tenths of a GV. Since the detector is located outside the atmosphere, the integral generation factor, $m^i(R, h_0)$ i.e., the result of the nuclear cascade process in the atmosphere, must be replaced by cascade multiplication in the spacecraft's surrounding material. For a spacecraft detector, in the limiting case, $m^i(R) \equiv 1$, but it is necessary to take into account the cascade multiplication of particles in the surrounding material, and the efficiency of any detector depends on the rigidity of the detected particles.